\documentclass[twocolumn,conference]{IEEEtran}

\IEEEoverridecommandlockouts

\usepackage{color}
\usepackage{graphicx}
\usepackage{amsmath}
\usepackage{amssymb}
\usepackage{algorithm}
\usepackage{algorithmic}
\usepackage{amsmath}
\usepackage{multirow}
\usepackage{booktabs}
\usepackage{array}
\usepackage{amsthm}
\usepackage{stfloats}
\usepackage{caption}
\usepackage{bm}
\usepackage{epstopdf}
\usepackage{gensymb} %\degree
\usepackage{subfigure} %后加
\usepackage{url} 
\usepackage{flushend}

\newcommand{\be}{\begin{equation}}
\newcommand{\ee}{\end{equation}}
\newcommand{\bea}{\begin{eqnarray}}
\newcommand{\eea}{\end{eqnarray}}
\newcommand{\ba}{\begin{array}}
\newcommand{\ea}{\end{array}}

\makeatletter

\newcommand{\Rmnum}[1]{\expandafter\@slowromancap\romannumeral #1@}
\makeatother
\newcommand{\RNum}[1]{\uppercase\expandafter{\romannumeral #1\relax}}

\title{End-to-End Learning for SLP-Based ISAC Systems
\thanks{$^{\ast}$ Corresponding author.}
\thanks{This work is supported in part by the National Natural Science Foundation of China (Grant No. 62371090 and 62071083), in part by Liaoning Applied Basic Research Program (Grant No. 2023JH2/101300201), and in part by Dalian Science and Technology Innovation Project (Grant No. 2022JJ12GX014).}}
\author{\IEEEauthorblockN{Yixian Zheng$^{\dag}$, Rang Liu$^{\ddag}$, Ming Li$^{\dag}$, and Qian Liu$^{\dag}$
\vspace{-0.0 cm} }
\IEEEauthorblockA{$^{\dag}$ Dalian University of Technology, Dalian, Liaoning 116024, China \\ E-mail: \texttt{zhengyixian@mail.dlut.edu.cn, \{mli, qianliu\} @dlut.edu.cn} }
\IEEEauthorblockA{$^{\ddag}$ University of California, Irvine, CA, USA \\ E-mail: \texttt{rang12@uci.edu} }
}

\begin{document}
\maketitle
\thispagestyle{empty}
\begin{abstract}
Integrated sensing and communication (ISAC) is an encouraging wireless technology which can simultaneously perform both radar and communication functionalities by sharing the same transmit waveform, spectral resource, and hardware platform. 
Recently emerged symbol-level precoding (SLP) technique exhibits advancement in ISAC systems by leveraging the waveform design degrees of freedom (DoFs) in both temporal and spatial domains. 
However, traditional SLP-based ISAC systems are designed in a modular paradigm, which potentially limits the overall performance of communication and radar sensing. 
The high complexity of existing SLP design algorithms is another issue that hurdles the practical deployment. 
To break through the bottleneck of these approaches, in this paper we propose an end-to-end approach to jointly design the SLP-based dual-functional transmitter and receivers of communication and radar sensing. 
In particular, we aim to utilize deep learning-based methods to minimize the symbol error rate (SER) of communication users, maximize the detection probability, and minimize the root mean square error (RMSE) of the target angle estimation. 
Multi-layer perceptron (MLP) networks and a long short term memory (LSTM) network are respectively applied to the transmitter, communication users and radar receiver. 
Simulation results verify the feasibility of the proposed deep-learning-based end-to-end optimization for ISAC systems and reveal the effectiveness of the proposed neural networks for the end-to-end design.
\end{abstract}


\maketitle
%\vspace{-0.3 cm}
\section{Introduction}
Next-generation wireless networks are universally regarded as a crucial driver for many exciting applications such as smart cities, smart healthcare, and autonomous driving. 
To meet their demands of both high-quality wireless communication and accurate sensing capability, integrated sensing and communication (ISAC) is emerging as one of the key enabling techniques for future wireless networks \cite{ISAC1}, \cite{ISAC11}.
In addition, symbol-level precoding (SLP) technique that can exploit the available degrees of freedom (DoFs) in both temporal and spatial domains is deemed as a promising candidate for ISAC systems, where the transmit waveform simultaneously carries information symbols and serves for radar sensing \cite{ISAC-SLP1}, \cite{ISAC-SLP2}.

Since radar sensing and communication functions are inherently conflicting requirements, transmit waveform design is crucial in the pursuit of better performance trade-offs for ISAC systems. 
Therefore, many researchers have investigated transmit waveform designs using various radar sensing and communication metrics, such as beampattern similarity, signal-to-interference-plus-noise ratio (SINR), achievable rate, etc. 
However, all of the above are intermediate performance metrics, as the system is designed as a modular paradigm. 
Although the traditional modular approach has helped us realize the present efficient and controllable systems, there are still some potential problems. 
For example, the local optimization of each module does not imply the end-to-end global optimization of the whole system \cite{introduction}, which points out a promising breakthrough for addressing performance bottlenecks in end-to-end research. 
In addition, the principle of SLP is to design nonlinear precoding for each transmitted symbol vector, resulting in extremely high complexity due to the huge number of precoders to be optimized during each channel coherent time. 
Therefore, deep learning-based end-to-end (E2E) optimization methods, which utilize powerful data learning capabilities, are a promising solution for the design of SLP-based ISAC systems.

Recently, deep learning-based approaches have been in full swing in the E2E research field. 
In \cite{SLP}, \cite{robust}, an E2E learning based approach is applied to SLP communication systems respectively for adaptive modulation and robust design for imperfect channel state information (CSI). 
Besides, E2E approach has also been applied to radar systems \cite{radar}, \cite{radar2}.
E2E learning for ISAC systems is proposed in \cite{E2E} for the first time.
However, the simple network structure and the block-level precoding (BLP) technique cannot provide satisfactory performance in more complex and practical ISAC applications. 
Furthermore, perfect CSI is necessary for the system, which is difficult to acquire in practical systems.

In this paper, we investigate the powerful E2E networks for the joint design of the dual-functional transmitter and receivers in the single-target and multi-user ISAC systems. 
In specific, the transmitter utilizes the SLP technique to generate a robust dual-function waveform, in order to fully exploit the DoFs in both temporal and spatial domains and combat the channel uncertainty when only limited priori angular information is available. 
The transmitter, communication receivers, and target detector are jointly designed as different multi-layer perceptron (MLP) networks respectively. 
For the scenario that the target is detected to be present, we utilize a long short term memory (LSTM) network to further estimate its angle, with a novel configuration for input dimension applied to better capture the angular information in the reflected echo signals. 
Moreover, a suitable combined loss function for all the decoding, detection, and estimation tasks is designed for optimization, aiming to minimize the symbol error rate (SER) of users, maximize the target detection probability, and minimize the root mean square error (RMSE) of the target angle estimation. 
Finally, simulation results are provided to confirm the feasibility of the proposed deep learning-based E2E optimization for ISAC systems and verify the effectiveness of the proposed neural networks in the E2E designs.

\section{System Model}
We consider an ISAC system, in which a dual-functional base station (BS) simultaneously serves multiple communication users and senses one potential target.
In particular, the BS is equipped with $N_\mathrm{t}$ transmit antennas and $N_\mathrm{r}$ receive antennas in uniform linear arrays (ULAs), which are respectively used for transmitting dual-functional signals and receiving echoes from the target.
Through properly optimizing transmit waveforms, the BS aims to communicate with $K$ single-antenna users and sense one point-like target.
Meanwhile, the BS detects the presence of the target within the area of interest and then further estimates its direction of arrival (DoA) by analyzing the received echo signals, and the communication users decode the received signals to extract desired messages. 

\subsection{SLP-based ISAC Transmitter}
The ISAC transmitter attempts to transfer message $m_k \in \mathcal{M}$ to the $k$-th user, $k=1,\ldots,K$, as well as sense one potential target. 
For simplicity, we denote the transferred message for $K$ users as $\mathbf{m} \triangleq \left[m_{1}, \ldots, m_{K}\right]^{T}$.
With the available channel state information (CSI) at the transmitter, the transmitted signal $\mathbf{x}$ can be obtained as a nonlinear precoding design $\mathbf{x}=\mathcal{F}(\mathbf{H},\mathbf{m})$ according to the principle of SLP, where $\mathcal{F}$ represents the mapping of symbol-level precoding, $\mathbf{H} \triangleq [\mathbf{h}_{\mathrm{c},1}, \ldots, \mathbf{h}_{\mathrm{c},K},\mathbf{H}_\text{r}]^{T}$, and $\mathbf{h}_{\mathrm{c},k} \in \mathbb{C}^{N_\text{t}}$ and $\mathbf{H}_\text{r} \in \mathbb{C}^{N_\text{t} \times N_\text{r}}$ respectively denote the channel vector between the transmitter and the $k$-th user and the channel matrix between the transmitter and the target.
However, it is difficult to acquire perfect CSI in practice. 
Therefore, in this paper we aim to design a SLP scheme only using the priori angular information $\Theta = [\theta_{\mathrm{min}},\theta_{\mathrm{max}},\vartheta_{1\mathrm{min}},\vartheta_{1\mathrm{max}},\ldots,\vartheta_{K\mathrm{min}},\vartheta_{K\mathrm{max}}]$, representing the potential target known to lie in a certain angle-of-arrival (AoA)  interval $\theta \sim \mathcal{U}(\theta_{\mathrm{min}},\theta_{\mathrm{max}})$, and the $k$-th user known to lie in $\vartheta_k \sim \mathcal{U}(\vartheta_{k\mathrm{min}},\vartheta_{k\mathrm{max}})$.
The non-linear mapping then can be expressed as
\begin{equation}
	\label{eq:transmitter}
	\mathbf{x}=\mathcal{F}(\Theta,\mathbf{m}), 
\end{equation} 
which should satisfy the power constraint: $\mathbb{E}\{\|\mathbf{x}\|^{2}\} \leq P$. 
For convenience, we denote the transmitted signal in the $n$-th time slot as $\mathbf{x}[n] \triangleq \left[x_{1}[n], \ldots, x_{N_{\mathrm{t}}}[n]\right]^{T}, n=1, \ldots, N$, where $N$ is the total number of time slots during one coherent processing interval (CPI). 
In order to achieve satisfactory multiuser communication and radar sensing performance,  the transmitted dual-functional signal $\mathbf{x}$ should be carefully designed.

\subsection{Single-target Sensing Model}
The dual-functional signal $\mathbf{x}$ is emitted towards the potential target, which will reflect the signal back to the BS.
Thus, the received echo signal in the $n$-th time slot can be written as \cite{channel}
\begin{equation}
	\label{e1}
	\mathbf{y}_{\mathrm{r}}[n]=\left\{\begin{array}{l}
		\mathcal{H}_{1}: G_{\mathrm{r}} {\alpha_\text{t}}\alpha_\text{d} \mathbf{a}_{\mathrm{r}}(\theta) \mathbf{a}_{\mathrm{t}}^{T}(\theta) \mathbf{x}[n]+\mathbf{z}_{\mathrm{r}}[n] \\
		\mathcal{H}_{0}: \mathbf{z}_{\mathrm{r}}[n]
	\end{array}\right.,
\end{equation}
where $\mathcal{H}_{1}$ and $\mathcal{H}_{0}$ represent the hypothesis of the received echo signals, $G_{\mathrm{r}}=\sqrt{N_{\mathrm{t}}N_{\mathrm{r}}}$ is the antenna array gain, $\alpha_\text{t}$ is the radar cross section (RCS) of the target, $\alpha_\text{d} = \alpha_0({d_{\mathrm{r},l}}/{d_{0}})^{-\gamma}$ is the distance-dependent path-loss at the distance $d_{\mathrm{r},l}$ with $\alpha_0$ being the path-loss at the reference distance $d_{0}$ and $\gamma$ the path-loss exponent, $\mathbf{z}_{\mathrm{r}}[n] \sim \mathcal{C N}(\mathbf{0}, {\sigma_\mathrm{r}}^2 \mathbf{I}_{N_\mathrm{r}})$ is the additive white Gaussian noise (AWGN), $\mathbf{a}_{\mathrm{t}}(\theta) \triangleq \frac{1}{\sqrt{N_{\mathrm{t}}}} \left[1, \ldots, e^{-{\jmath 2 \pi}/{\lambda}\left(N_{\mathrm{t}}-1\right) d \sin \theta}\right]^{T}$ is the transmit steering vector for the potential target at the direction of $\theta$, with $d={\lambda}/{2}$ denoting the antenna spacing and $\lambda$ the wavelength, and $\mathbf{a}_{\mathrm{r}}(\theta)$ is the receiving steering vector defined similarly. 
To perform radar sensing functionality, the received echo signals during $N$ time slots are collected at the BS for further processing. 
For simplicity, we stack the received echo signals into a matrix as $\mathbf{Y}_\mathrm{r} \triangleq \left[\mathbf{y}_\mathrm{r}[1], \ldots, \mathbf{y}_\mathrm{r}[N]\right] \in \mathbb{C}^{N_\mathrm{r} \times N}$, which is utilized to detect the presence of the potential target and further estimate its accurate DoA.
Particularly, the presence of the target is distinguished by the presence detector as $\widehat{t}=\mathcal{D}(\mathbf{Y}_\text{r},\theta_\text{min},\theta_\text{max})$, and then the DoA of the target is estimated as $\widehat{\theta}=\mathcal{A}(\mathbf{Y}_\text{r})$. 
In the sequel, the detection probability is $P_\text{d} = \text{Pr}(\widehat{t}=1 | t=1)$, and the RMSE of target DoA estimation can be calculated as $\text{RMSE}=\sqrt{\mathbb{E}\{|\widehat{\theta}-\theta|^2\}}$.

\subsection{Multi-user Communication Model}
After propagating through the downlink communication channel, the signal received at the $k$-th user in the $n$-th time slot is given by
\begin{equation}
	\label{e2}
	y_{\mathrm{c},k}[n]=G_{\mathrm{c}} \sqrt{\beta_\mathrm{d}}  \mathbf{a}_{\mathrm{t}}^{T}(\vartheta_k) \mathbf{x}[n]+z_{\mathrm{c},k}[n],
\end{equation}
where $G_{\mathrm{c}}=\sqrt{N_{\mathrm{t}}}$ is the antenna array gain, $\beta_\mathrm{d}$ is the path-loss coefficient, $z_{\mathrm{c},k}[n] \sim \mathcal{C N}(0,{\sigma_\mathrm{c}}^2)$ is the AWGN, and $\vartheta_k$ is the DoA of the $k$-th user. 
Given received noise-corrupted signals, the users attempt to decode their desired messages through appropriate methods and retrieve the message as $\widehat{m}_k = \mathcal{C}_k(y_{\text{c},k})$. Thus, the average SER of the users is $\text{SER}_\text{a} = \frac{1}{K}\sum_{k=1}^K\mathbb{E}\{\text{Pr}(\widehat{m}_k\neq m_k)\}$.

\subsection{Problem Formulation}

In this paper, we propose to jointly design the symbol-level precoding $\mathcal{F}$ at the transmitter, the decoder $\mathcal{C}_k$ at the $k$-th communication user, $\forall k$, the target presence detector $\mathcal{D}$ and the angle estimator $\mathcal{A}$, in order to minimize the average SER of multi-user communications $\text{SER}_\text{a}$, maximize the detection probability $P_\text{d}$ and minimize the RMSE of angular estimation of the target as well as to satisfy the transmit power budget $P$.
Thus, the joint transmitter and receiver design problem can be formulated as
\begin{subequations}
	\label{eq:problem}
	\begin{align}\label{eq:problem a}
		\min\limits_{\mathcal{F}, \mathcal{C}_k, \mathcal{D}, \mathcal{A}, \forall k}~~&
		\text{SER}_\text{a} - \mu_1 P_\text{d} + \mu_2 \text{RMSE} \\
		\label{eq:problem b}
		\textrm{s.t.}~~& \mathbf{x}=\mathcal{F}(\Theta,\mathbf{m}),\\
		\label{eq:problem c}
		~~& \widehat{m}_k=\mathcal{C}_k(y_{\text{c},k}),\\
		\label{eq:problem d}
		~~& \widehat{t}=\mathcal{D}(\mathbf{Y}_\text{r},\theta_{\mathrm{min}},\theta_{\mathrm{max}}),\\
		~~& \widehat{\theta}=\mathcal{A}(\mathbf{Y}_\text{r}),\\
		\label{eq:problem d}
		~~& \|\mathbf{x}\|^{2} \le P,
	\end{align}
\end{subequations}
where $\mu_1$ and $\mu_2$ are hyper-parameters to adjust the magnitudes of these three performance indicators and realize the performance trade-off among different tasks. 
It is obvious that the joint transmitter and receivers design in problem (\ref{eq:problem}) cannot be modeled as traditional optimization problems and then solved using typical optimization algorithms. 
In order to overcome these difficulties, in the next section we propose a deep learning-based end-to-end design approach for the considered ISAC systems.

%\vspace{-0.2 cm}
\section{End-to-End Learning for ISAC}

\begin{figure*}
	\centering
	%\vspace{-0.5 cm}
	\includegraphics[width= 5.5 in]{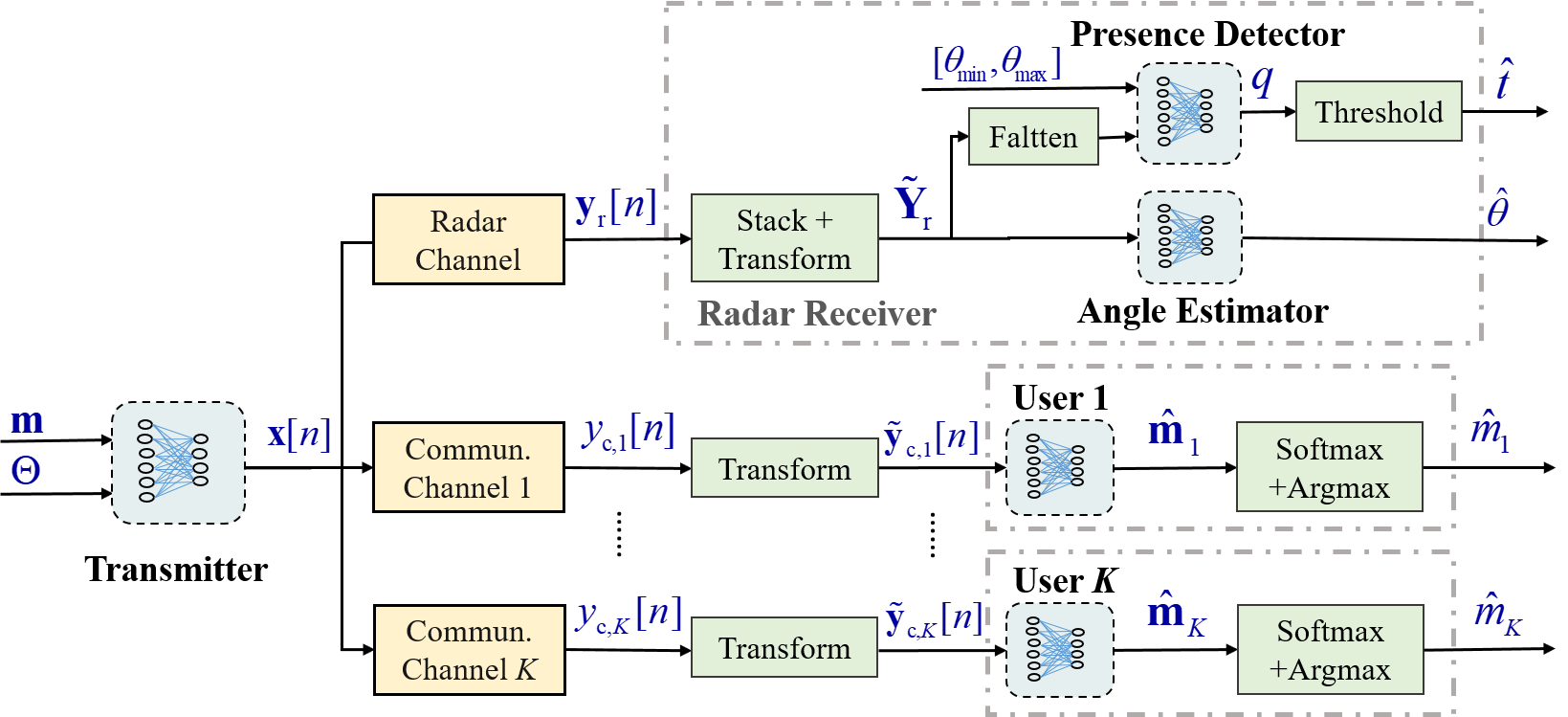}
	%\vspace{-0.5 cm}
	\caption{The block diagram of the considered ISAC system model.} %e2e
	\label{fig:block diagram}
	%\vspace{-0.3 cm}
\end{figure*}

In this section, we propose a deep learning-based end-to-end approach to realize ISAC, whose block diagram is illustrated in Fig. \ref{fig:block diagram}.
The blocks in blue network sketches are separately implemented by deep neural networks (NNs).
The blocks highlighted in green indicate the pre-processing for input data. 
The details are as follows.

\subsection{Transmitter}
\subsubsection{SLP-based transmitter}
\begin{figure}[h]
	\centering
	%\vspace{-0.5 cm}
	\includegraphics[width= 0.55\linewidth]{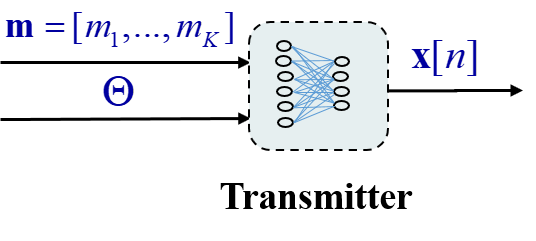}
	%\vspace{-0.5 cm}
	\caption{Symbol-level precoding.}
	\label{fig:transmitter1}
	%\vspace{-0.3 cm}
\end{figure}

As shown in Fig. \ref{fig:transmitter1}, based on the priori information for the angular range of communication users and potential target $\Theta = [\theta_{\mathrm{min}},\theta_{\mathrm{max}},\vartheta_{1\mathrm{min}},\vartheta_{1\mathrm{max}},\ldots,\vartheta_{K\mathrm{min}},\vartheta_{K\mathrm{max}}]$ and the transmitted message $\mathbf{m}$, the transmitter designs the symbol-level precoded signal $\mathbf{x}$, expected to realize satisfactory communication and radar sensing performance and be robust to channel uncertainty.
The transmitter is designed as an MLP network. The detailed size of layers and the corresponding activation functions for the encoder and beamformer are shown in Table \ref{table: mlp}.
A normalization layer is applied to meet the power constraint after the network, which can be expressed as $\mathbb{E}\{\|\mathbf{x}\|^{2}\} = P$. 

\begin{figure}[h]
	\centering
	%\vspace{-0.5 cm}
	\includegraphics[width= 0.85\linewidth]{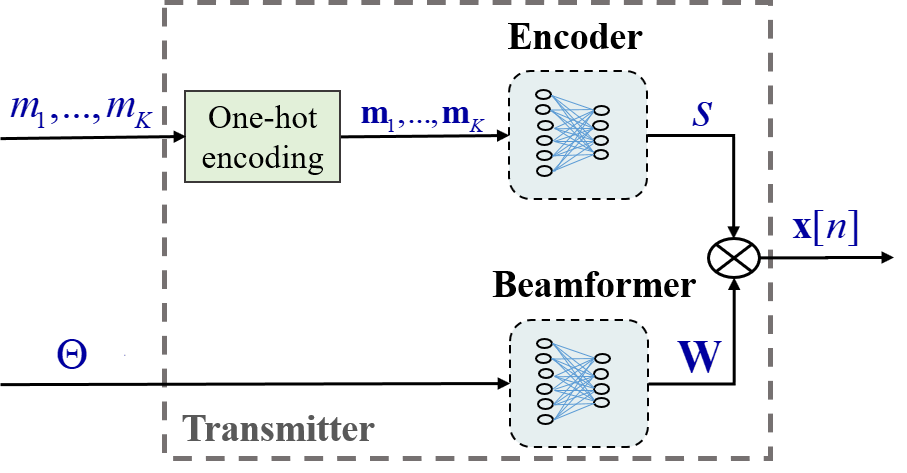}
	%\vspace{-0.5 cm}
	\caption{Block-level precoding.}
	\label{fig:transmitter2}
	%\vspace{-0.3 cm}
\end{figure}

\subsubsection{BLP-based Transmitter}
In order to demonstrate the advantage of SLP-based transmitter, we also design the typical BLP-based transmitter consisting of an encoder and a beamformer as a comparison, shown in Fig. \ref{fig:transmitter2}.
As typically implemented in end-to-end communication systems, we first utilize one-hot encoding to convert the message $m_k \in \mathcal{M}$ into the vector $\mathbf{m}_k \in \mathbb{R}^{|\mathcal{M}|}$. 
Given the one-hot encoded vector $\mathbf{m}_k$, the encoder generates the modulation symbol $s_k$ for information transmission. 
Based on the priori angular information $\Theta$, the beamformer designs the block-level precoding matrix $\mathbf{W} \triangleq [\mathbf{w}_1, \ldots, \mathbf{w}_K] \in \mathbb{C}^{N_\mathrm{t} \times K}$ in order to focus the beam towards the direction of the targets and users and transmit the message $m_k$ to the corresponding user $k$.
Thus, the transmitted signal can be expressed as
\begin{equation}
	\label{e1}
	\mathbf{x}=\mathbf{W}\mathbf{s},
\end{equation}
where $\mathbf{s} \triangleq [s_1, \ldots, s_K]^T\in \mathbb{C}^{K}$.
The encoder and beamformer are both established as 5-layer MLP networks, of which the number of neurons at different layers is [$|\mathcal{M}|,N_\text{t},N_\text{t},2N_\text{t},2$] and [$8,N_\text{t},4N_\text{t},8N_\text{t},2N_\text{t}K$], respectively. 

\begin{table}[h]
	\centering
	\caption{Detailed MLP architecture. Assume $N_\mathrm{t} = N_\mathrm{r}$.}
	\label{table: mlp}
	\begin{tabular}{|c|c|c|c|}
		\hline
		Network                            & Input & Output & Activation function \\ \hline
		\multirow{5}{*}{Transmitter}           & $8+K$     & $2N_\mathrm{t}$      & ReLU                \\ \cline{2-4} 
		& $2N_\mathrm{t}$     & $4N_\mathrm{t}$      & ReLU                \\ \cline{2-4} 
		& $4N_\mathrm{t}$     & $8N_\mathrm{t}$     & ReLU                \\ \cline{2-4} 
		& $8N_\mathrm{t}$    & $4N_\mathrm{t}$      & ReLU
		\\ \cline{2-4} 
		& $4N_\mathrm{t}$    & $2N_\mathrm{t}$      & Linear              \\ \hline
		\multirow{4}{*}{Presence Detector} & $2NN_\mathrm{t}+2$ & $NN_\mathrm{t}$    & ReLU                \\ \cline{2-4} 
		& $NN_\mathrm{t}$   & $N$      & ReLU                \\ \cline{2-4} 
		& $N$     & $N_\mathrm{t}$      & ReLU                \\ \cline{2-4} 
		& $N_\mathrm{t}$     & $1$      & Sigmoid             \\ \hline
		\multirow{5}{*}{ Commun. Receiver}           & $2$     & $N_\mathrm{t}$      & ReLU                \\ \cline{2-4} 
		& $N_\mathrm{t}$     & $2N_\mathrm{t}$    & ReLU                \\ \cline{2-4} 
		& $2N_\mathrm{t}$   & $2N_\mathrm{t}$    & ReLU                \\ \cline{2-4} 
		& $2N_\mathrm{t}$   & $2N_\mathrm{t}$    & ReLU                \\ \cline{2-4} 
		& $2N_\mathrm{t}$   & $|\mathcal{M}|$      & Linear              \\ \hline
	\end{tabular}
\end{table}

\subsection{Communication Receivers}
The decoder at communication receivers is also realized by MLP networks. 
In particular, at the $k$-th user, the received noise-corrupted signal $y_{\mathrm{c},k}[n]$ is first transformed into a real-valued form  $\widetilde{\mathbf{y}}_{\mathrm{c},k}[n] \in \mathbb{R}^2$ as 
\begin{equation}
	\label{eq:y-prepocessing}
	\widetilde{\mathbf{y}}_{\mathrm{c},k}[n] = [\mathrm{Re}(y_{\mathrm{c},k}), ~~\mathrm{Im}(y_{\mathrm{c},k})]^T.
\end{equation}
This is because it is very difficult to handle complex data with existing popular deep learning libraries.
Through the message decoder, we obtain an $|\mathcal{M}|$-dimensional probability vector $\widehat{\mathbf{m}}_k \in [0,1]^{|\mathcal{M}|}$, the $i$-th element of which represents the probability that the decoded result is the $i$-th message in $\mathcal{M}$, $i = 1,\ldots,|\mathcal{M}|$. 
Then, the index of the element which has the largest probability is adopted as the decoded message. 
In order to lower communication SER, the decoder is designed to provide a more accurate decision rule. 
The detailed size of layers and the corresponding activation functions for the communication receivers are shown in Table \ref{table: mlp}.

\subsection{Target Presence Detector}
For the target detection functionality, we first flatten the stacked signals $\mathbf{Y}_\mathrm{r} \in \mathbb{C}^{N_\mathrm{r} \times N}$ during $N$ time slots into a vector $\widetilde{\mathbf{y}}_\mathrm{r} \in \mathbb{C}^{N N_\mathrm{r}}$ and concatenate the angular priori information $[\theta_{\mathrm{min}}, \theta_{\mathrm{max}}]$ of the target, then utilize an MLP network to process $[\widetilde{\mathbf{y}}_\mathrm{r}, \theta_{\mathrm{min}}, \theta_{\mathrm{max}}] \in \mathbb{C}^{N N_\mathrm{r}+2}$ and obtain a probability indicator $q \in [0,1]$.
With the hyper-parameter $\overline{q}$, which represents the threshold to judge the presence of the target, the target is deemed as present ($\widehat{t}=1$) when $q \geq \overline{q}$ and absent ($\widehat{t}=0$) when $q < \overline{q}$, respectively. 
The detailed sizes of layers and the corresponding activation functions for the presence detector are shown in Table \ref{table: mlp}.

\subsection{Target Angle Estimator}
With the received signals from multiple antennas and during multiple time slots, the angle estimator is designed as an LSTM network to estimate the DoA of the potential target. 
The proposed neural network is composed of the input pre-processing, the LSTM cell, and the fully-connected layer, whose details are presented below.
 
%\subsubsection{Input Pre-Processing}
Given the received echo signal matrix $\mathbf{Y}_\mathrm{r}$ with dimension $N_\mathrm{r} \times N$, we divide the complex-valued input into the real and imaginary parts, and concatenate them into the matrix $\widetilde{\mathbf{Y}}_\mathrm{r} \in \mathbb{R}^{2N_\mathrm{r} \times N}$ as 
\begin{equation}
	\label{eq:angle-preprocessing}
	\widetilde{\mathbf{Y}}_\mathrm{r}=\left[\begin{array}{l}
		\mathrm{Re}(\mathbf{Y}_\mathrm{r}) \\
		\mathrm{Im}(\mathbf{Y}_\mathrm{r})
	\end{array}\right],
\end{equation}
the $n$-th column of which is denoted as $\widetilde{\mathbf{y}}_\mathrm{r}[n]$. It is obvious that the vector $\widetilde{\mathbf{y}}_\mathrm{r}[n] \in \mathbb{R}^{2N_\mathrm{r}}$ is the concatenated vector of the real and imaginary parts of $\mathbf{y}_\mathrm{r}[n]$.

In addition, it is noted that the input to the LSTM network has three dimensions, i.e., the sample number, the timeline, and the feature number, which respectively represent the training batch size, data at different moments, and features of the input data at the moment. 
It is obvious that the relationship in time delays of the received signals at different antennas contributes to the DoA estimation performance, while the temporal relationship between the signals received at different time slots has a very marginal impact on the estimation performance. 
Therefore, we treat the signals received by different antennas as the timeline dimension of the LSTM network and treat the signals received at different time slots as the feature number dimension.

With the aid of the mechanism of memory cells, the LSTM network can capture both the short-term and long-term dependencies in the input sequences along the dimension of antennas, which will facilitate DoA estimation.

Finally, we use a fully-connected layer to transform the output of the last LSTM cell, and then employ the $\tanh$ activation function and scaling operation to obtain the final output $\widehat{\theta}$.

\subsection{Loss Functions}

In order to achieve satisfactory communication and radar sensing performance and flexible performance trade-off, we utilize a combined loss function to train the proposed end-to-end neural network.

\subsubsection{Detection Loss}
We use the binary cross-entropy (BCE) between the estimated presence probability $q$ and the presence label $t$ as the loss function, which is expressed as
\begin{equation}
	\label{e: detection loss}
	\mathcal{L}_1=-\mathbb{E}[t \log (q)+(1-t) \log (1-q)].
\end{equation}

\subsubsection{Angular Estimation}

If the target is present, we apply the mean squared error (MSE) between the estimated angle $\widehat{\theta}$ and actual angle $\theta$ as the loss function, which is expressed as
\begin{equation}
	\label{e: eastimation loss}
	\mathcal{L}_2=\mathbb{E}\big[|\widehat{\theta}-\theta|^2\big].
\end{equation}

\subsubsection{Communication Loss}
The categorical cross-entropy (CCE) between the decoded vector $\widehat{\mathbf{m}}_{k}$ and the actual message $\mathbf{m}$ is utilized as the loss function for the $k$-th user.
The overall loss function for all the $K$ users is expressed as
\begin{equation}
	\label{e: commun loss}
	\mathcal{L}_3=-\sum_{k=1}^{K} \mathbb{E}\left[\sum_{i=1}^{\mid\mathcal{M} \mid} \mathbf{m}_k(i) \log \left(\widehat{\mathbf{m}}_{k}(i)\right)\right].
\end{equation}

\subsubsection{Overall ISAC Loss}
Based on the above separate loss functions developed for the presence detector, angular estimation and communication receivers, we propose to use the linear combination of those individual ones as the joint loss function, which is expressed as
\begin{equation}
	\label{e: ISAC loss}
	\mathcal{L}_{\text {ISAC}}=\omega _2 [(1-\omega_1) \mathcal{L}_1+ \omega _1 \mathcal{L}_2] + (1-\omega _2) \mathcal{L}_3,
\end{equation}
where $\omega_1 \in [0,1]$ and $\omega_2 \in [0,1]$ are hyper-parameters. 
Moreover, $\omega_1$ is used to balance the gap in orders of magnitude between $\mathcal{L}_1$ and $\mathcal{L}_2$, and $\omega_2$ is used to adjust the performance trade-off between communication and radar sensing. 

%\vspace{-0.3 cm}
\section{Simulation Results}
In this section, we evaluate the performance of the proposed end-to-end learning approach for SLP-based ISAC systems.
We assume that the numbers of transmit antennas and receive antennas are the same, i.e., $N_\mathrm{t} = N_\mathrm{r}$.
The target is located at a distance of $d_{{\mathrm{r},l}}$ in the direction of $\theta$ with respect to the transmitter, where $d_{{\mathrm{r},l}} \sim \mathcal{C N}(d_\mathrm{r}, {\sigma_\mathrm{1}}^2)$ with $d_\mathrm{r}=10$m, ${\sigma_\mathrm{1}}^2=1$, $\theta \sim \mathcal{U}\left[-10, 10\right]\degree$, and the radar cross section $\alpha_\text{t} = 1$.
The $K=3$ users are located at the distance of $d_{{\mathrm{c},l}} \sim \mathcal{C N}(d_\mathrm{c}, {\sigma_\mathrm{2}}^2)$ with $d_\mathrm{c}=150$m, ${\sigma_\mathrm{2}}^2=1$ and in the direction of $\vartheta_1 \sim \mathcal{U}\left[50, 70\right]\degree$, $\vartheta_2 \sim \mathcal{U}\left[-75, -60\right]\degree$, $\vartheta_3 \sim \mathcal{U}\left[-45, -30\right]\degree$, respectively. 
The path loss at the reference distance $d_0 = 1$m is $\alpha_0 =\beta_0 = -30$dB, with the path loss exponent $\gamma = 2.2$.
The noise power is set as ${\sigma_\mathrm{c}}^2 ={\sigma_\mathrm{r}}^2 = -70$dBmW.
\begin{figure}[!t]
	\centering
	%\vspace{-0.5 cm}
	\includegraphics[width= 0.92\linewidth]{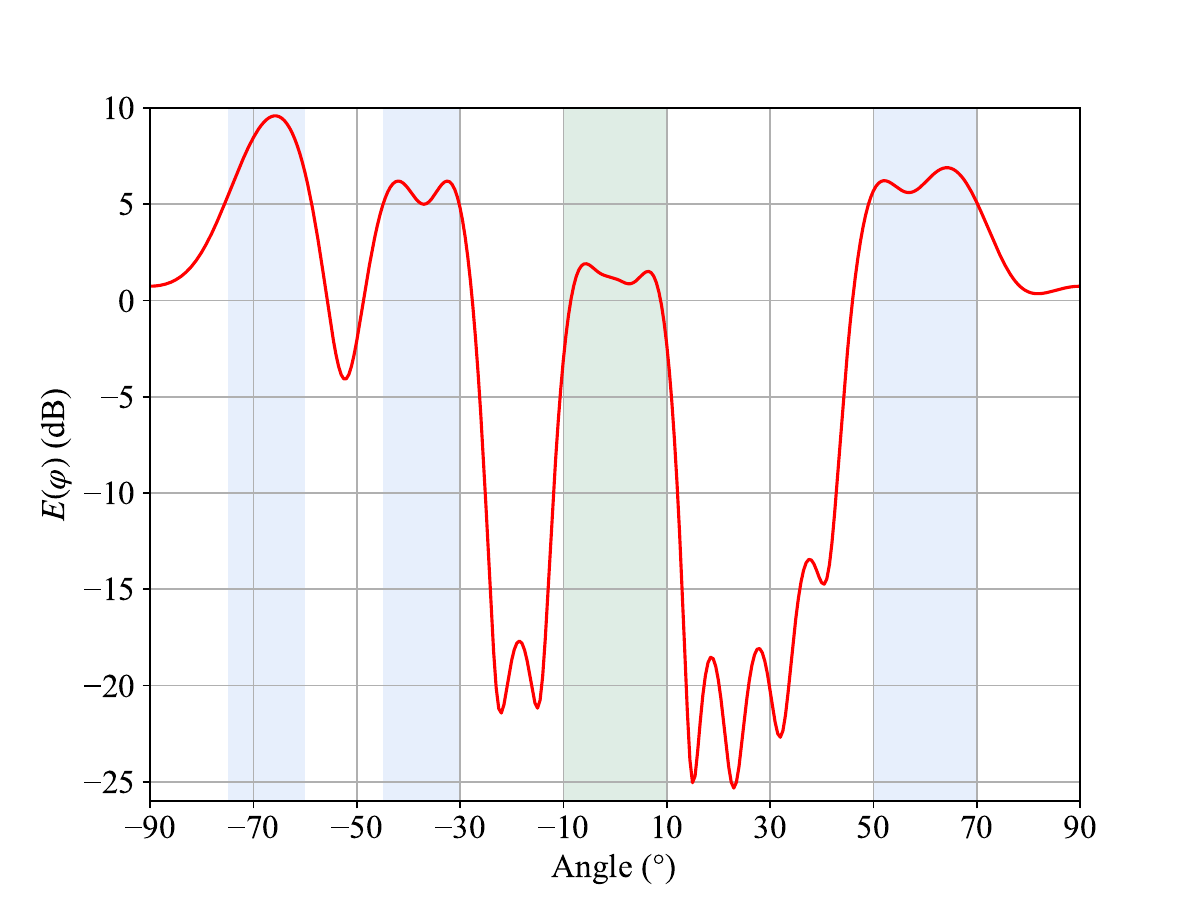}
	%\vspace{-0.5 cm}
	\caption{Transmit beampattern ($N_\mathrm{t} = 16$). }
	\label{fig:beampattern}
	%\vspace{-0.3 cm}
\end{figure}

We generate $10^6$ Monte Carlo channel realizations in establishing the training set, and $5 \times 10^5$ realizations in the testing set.
In addition, an equal number of cases of target presence and absence are generated in the data set.
The transmit power in the training stage is set as $P = 10$dBmW.
The hyper-parameters used in the ISAC loss function (\ref{e: ISAC loss}) are $\omega_1 = 0.05$ and $\omega_2 = 0.3$.
Given the overall ISAC loss in (\ref{e: ISAC loss}), we jointly train the end-to-end neural network in an unsupervised manner.
The Adam optimizer is used for training the networks with a learning rate of $10^{-3}$ and the batch size being 1000.

% M = 4, 8, 16
% $N_\mathrm{t} = N_\mathrm{r} = 16/32/64$
We first plot the transmit beampattern (the signal power in different directions) in Fig. \ref{fig:beampattern}, in which the green area is the possible angular area of the target, and the blue ones indicate the possible angular areas of the communication users.
We observe that the beams are focused on the possible angular regions of the target and the communication users, which offers visual and preliminary confirmation of the effectiveness of precoding.

\begin{figure}[!t]
	\centering
	%\vspace{-0.5 cm}
	\includegraphics[width= 0.92\linewidth]{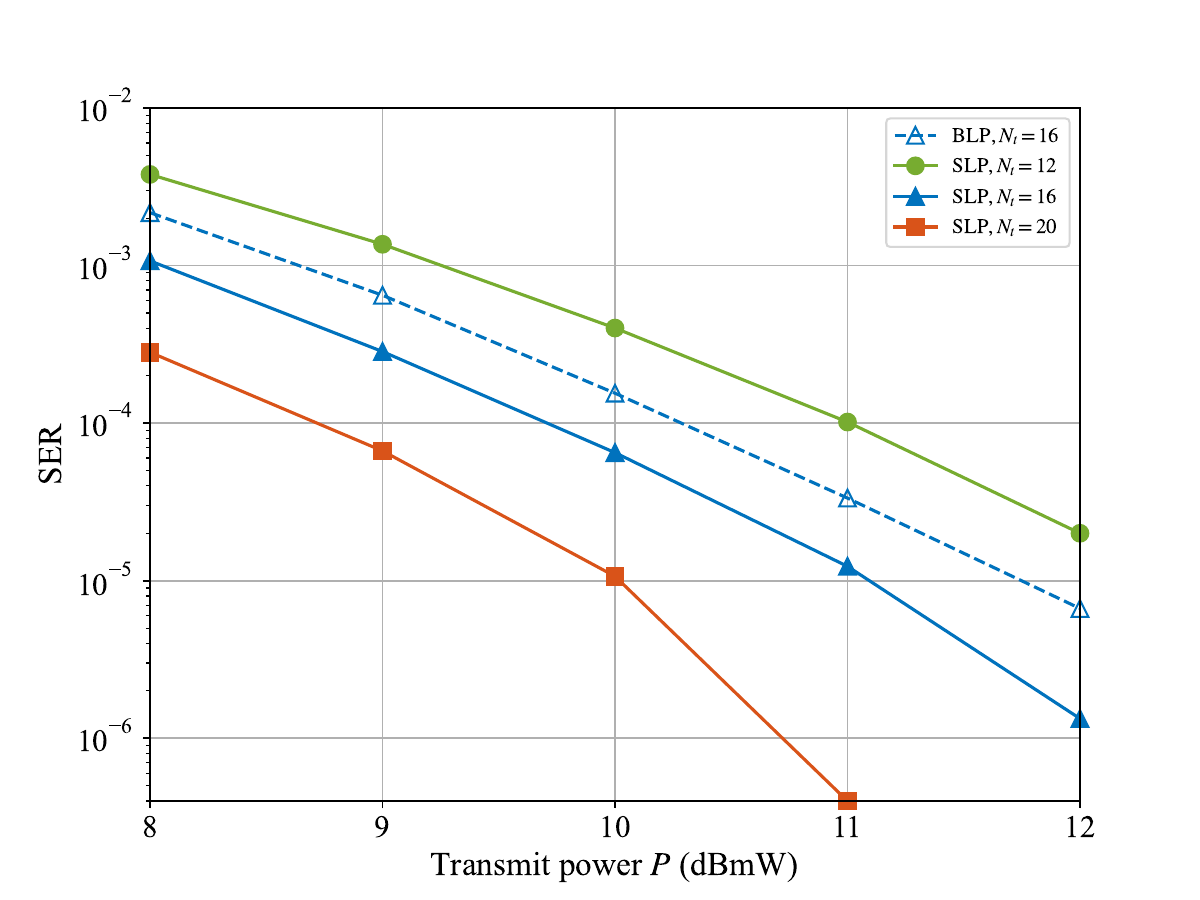}
	%\vspace{-0.5 cm}
	\caption{SER versus transmit power $P$.}
	\label{fig:SER}
	%\vspace{-0.3 cm}
\end{figure}

In Fig. \ref{fig:SER}, we show the average SER of the communication users versus the transmit power budget $P$. 
It is obvious that the communication receiver achieves a quite satisfactory SER performance within the reasonable transmit power range.
Furthermore, the SLP-based transmitter outperforms the BLP-based transmitter with about a 58$\%$ decrease in SER when $P=10$dBmW. 
In addition, it is shown that more transmit or receive antennas, which also means more neurons in the network, can bring notable performance improvements.
However, larger time and memory overheads are also caused by the increase of $N_\text{t}$. For example, for each epoch, it requires 383 seconds for $N_\text{t}=N_\text{r}=12$, 534 seconds for $N_\text{t}=N_\text{r}=16$, and 678 seconds for $N_\text{t}=N_\text{r}=20$ (about 1:1.39:1.77).
We only compare our proposed approach with BLP-based approach because, to the best of our knowledge, there is no research on robust SLP design for ISAC systems when only limited priori angular information is available.

\begin{figure}[!t]
	\centering
	%\vspace{-0.5 cm}
	\includegraphics[width= 0.92\linewidth]{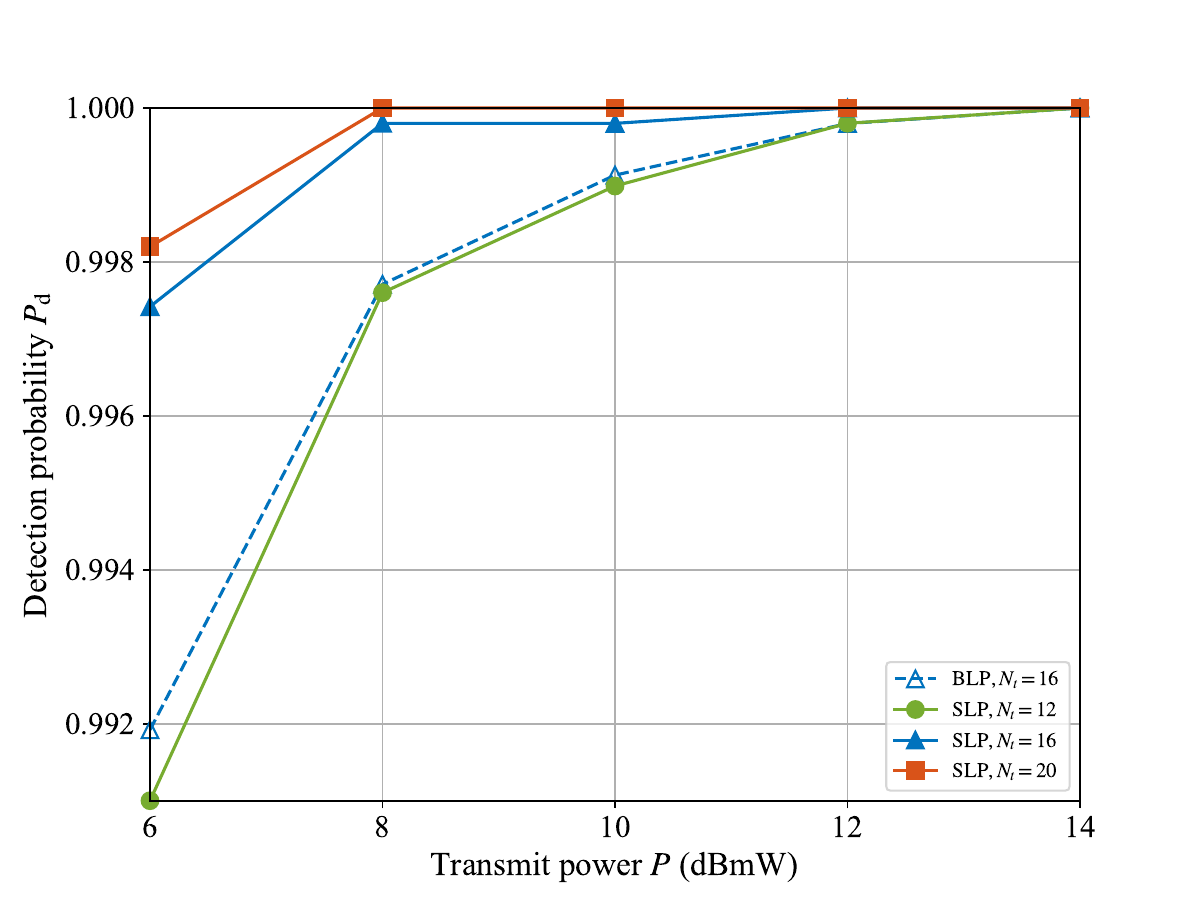}
	%\vspace{-0.5 cm}
	\caption{Detection probability $P_\mathrm{d}$ versus transmit power $P$.}
	\label{fig:Pd}
	%\vspace{-0.3 cm}
\end{figure}

In Fig. \ref{fig:Pd}, the detection probability $P_\mathrm{d}$ versus the transmit power $P$ is illustrated, from which satisfactory target detection performance can be observed. 
We also see an 80$\%$ decrease in $1-P_\mathrm{d}$ achieved by the proposed SLP scheme compared with the typical BLP scheme when $P=10$ dBmW, which indicates that SLP provides more DoFs for improving radar sensing performance.
Additionally, notable performance improvements can be obtained by increasing the number of antennas and neurons in the network, similar to Fig. \ref{fig:SER}.

 \begin{figure}[!t]
	\centering
	%\vspace{-0.5 cm}
	\includegraphics[width= 0.92\linewidth]{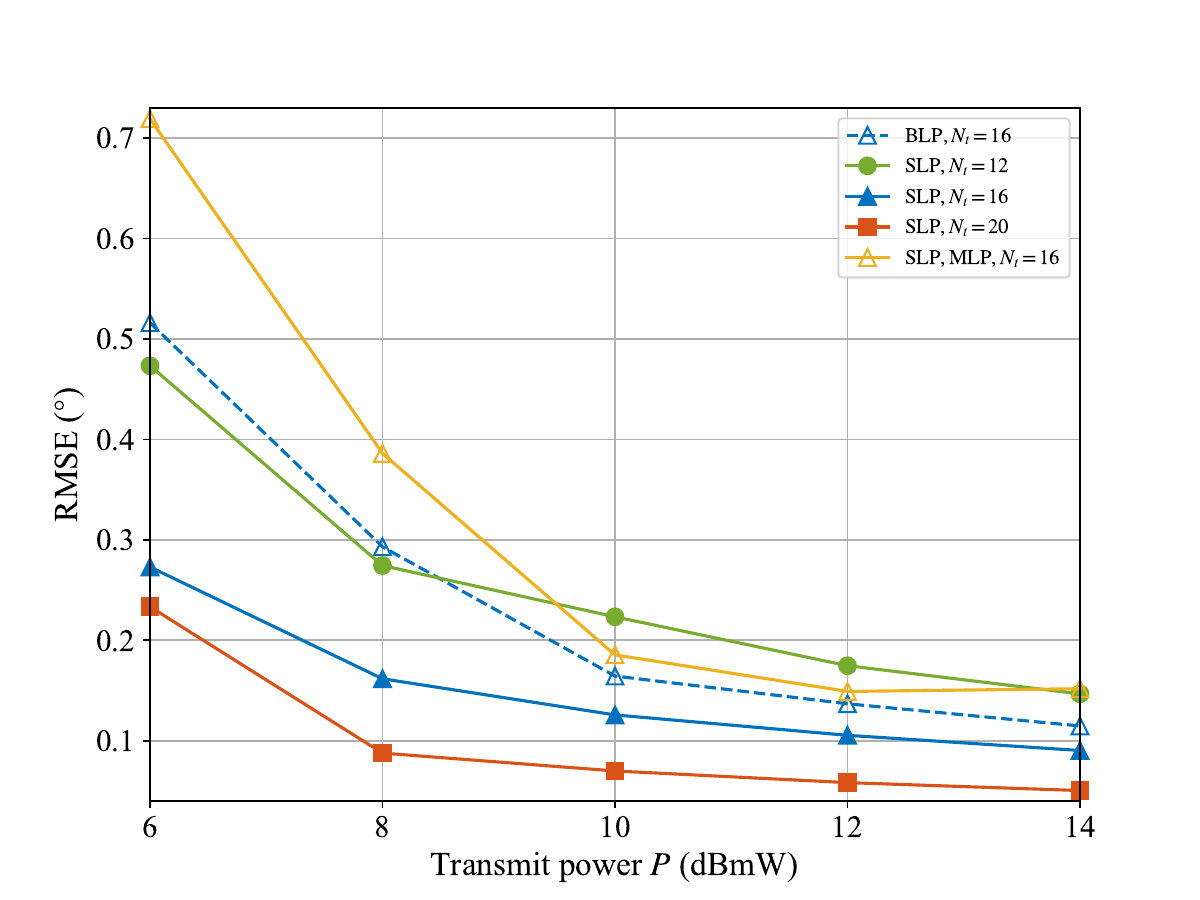}
	%\vspace{-0.5 cm}
	\caption{RMSE of angle estimation versus transmit power $P$.}
	\label{fig:RMSE}
	%\vspace{-0.3 cm}
\end{figure}

In Fig. \ref{fig:RMSE}, we show the RMSE of angular estimation versus the transmit power $P$ for the scenario that the target is detected to be present. 
A notable 23.5$\%$ decrease in RMSE is achieved with the aid of SLP-based transmitter in comparison with the typical BLP scheme when $P=10$ dBmW.
Similarly in Fig. \ref{fig:SER} and \ref{fig:Pd}, performance gain due to the increase of the number of antennas and neurons in the network can also be observed.
For further comparison, we also include the approach that utilizes an MLP network for angular estimation, of which the number of neurons at different layers are [$2NN_\text{t},NN_\text{t},N,N_\text{t},1$]. 
Even if we vary the transmit power $P \sim \mathcal{U}(6,14)$ through training this time to enable the network to adapt to different transmit power, satisfactory angular estimation performance is still only guaranteed within a small range of power.
It takes about 400 seconds on each training epoch, about 75\% and 104\% of the time budget of the LSTM network with $N_\text{t}=16$ and $12$ respectively. 
However, it performs much worse than the LSTM network with the same number of antennas, especially for the smaller transmit power budget, and it even performs worse than the LSTM network with fewer antennas and less time budget in the low transmit power range and only performs slightly better in the high transmit power range.
This is because that the MLP network is not well suited to the angular estimation task and may suffer from network overfitting easily. 
On the contrary, since the LSTM network fully exploits the reflected echo signals during one CPI and utilizes the mechanism of memory cells to solve the problems of gradient vanishing and gradient explosion, the angle-dependent correlations in the data can be more efficiently extracted and processed with longer-term effective memory.

%\vspace{-0.3 cm}
\section{Conclusions}

In this paper, we proposed an end-to-end learning approach to jointly optimize the SLP-based ISAC transmitter, the target detection and the angle estimation at the radar receiver, and the signal decoding at communication receivers, which is robust to channel uncertainty.
We established the end-to-end neural networks consisting of multi-layer perceptron (MLP) networks and a long short term memory (LSTM) network. 
Then, we proposed a combined loss function and trained the overall neural network in an unsupervised manner. 
Simulation results validated the feasibility and effectiveness of the proposed deep learning-based end-to-end approach for the considered SLP-based ISAC system.

\end{document}